\documentclass[showpacs,aps,prd,floatfix,amsmath,amssymb,nofootinbib]{revtex4-2}
\usepackage[utf8x]{inputenc}
\usepackage{graphicx}
\usepackage{amsmath}
\usepackage{array}
\usepackage{slashed}
\usepackage[makeroom]{cancel}
\usepackage{bbding}
\usepackage{xcolor}
\usepackage{multirow}
\usepackage[colorinlistoftodos]{todonotes}
\usepackage[colorlinks=true, allcolors=blue]{hyperref}
\definecolor{darkred}{rgb}{0.6,0,0}
\usepackage{comment}
\newcommand{\rlq}{R^{1/3}}
\newcommand{\slq}{S^{1/3}}
\begin{document}
\title{Neutrino magnetic moment in the doublet-singlet Leptoquark model}
\author{R. S\'anchez-V\'elez}
\email[Corresponding author: ]{rsanchezve@ipn.mx}
\affiliation{Departamento de F\'isica, Centro de Investigaci\'on y de Estudios Avanzados del IPN 
Apdo. Postal 14-740 07000 Ciudad de M\'exico, M\'exico}

\begin{abstract}
The transition magnetic moment for Majorana neutrinos is studied in a simple extension of the Standard Model. This extension incorporates two scalar Leptoquarks $S_1$ and $\widetilde{R}_2$ with quantum numbers $(\bar{3},1,1/3)$ and $(3,2,1/6)$ respectively. It is found that these Leptoquarks generate a sizable transition magnetic moment, particularly when the quark bottom is running in the loop. For our analysis of the parameter space, we include the latest measurement of the muon magnetic moment and combine it with the experimental constraint on the branching ratio Br$(\tau \to \mu \gamma)$. We found that, despite the recent agreement on the $(g-2)_\mu$ value, large values for Leptoquark Yukawa couplings are allowed due to a degeneracy in the parameters. We determine that the restrictions derived from our analysis are consistent with the most recent experimental limits reported by the XENONnT and LUX-ZEPLIN collaborations. This conclusion is based on our evaluation of the transition magnetic moment from muon neutrino to tau neutrino, focusing on the allowed region for the Leptoquark Yukawa couplings.
\end{abstract}
\maketitle

\section{Introduction}
\label{introduction}

In the Standard Model (SM), neutrinos are considered massless particles. However, it is possible to generate a neutrino magnetic moment, $\mu_{\nu_{\alpha \beta}}$ (diagonal $\alpha = \beta$ and transition $\alpha \neq \beta$), by adding right-handed neutrinos into the SM. The calculation of the diagonal magnetic moment for a Dirac neutrino yields to~\cite{Fujikawa:1980yx}
\begin{equation}\label{eq:NeutrinoMagnetic}
\mu_{\nu}=\frac{3 m_e G_F}{4\sqrt{2} \pi^2}m_\nu \mu_B\approx 3.2\times 10^{-19}\left(\frac{m_\nu}{eV}\right)\mu_B,
\end{equation}
where $m_\nu$ and $m_e$ are the neutrino and electron masses, respectively, $G_F$ is the Fermi constant and $\mu_B$ is the Bohr magneton used as a conventional unit. In such a minimal extension of the SM, the extremely small mass of neutrinos results in a magnetic moment that is far beyond experimental capabilities. Nevertheless, larger values can be achieved in many other frameworks beyond the minimally-extended SM, reaching values of the order of $10^{-12}\mu_B$ or even $10^{-10}\mu_B$~\cite{Aboubrahim:2013yfa,Lindner:2017uvt}. Examples of these frameworks include models with left-right symmetry~\cite{Czakon:1998rf,Boyarkin:2014oza}, R-parity-violating supersymmetry~\cite{Gozdz:2012xw}, large extra dimensions~\cite{Mohapatra:2004ce}, and non-standard neutrino interactions~\cite{Xu:2019dxe}. On the other hand, laboratory limits on the neutrino magnetic moment are established through neutrino(antineutrino)-electron scattering at low energies. The GEMMA collaboration has provided the upper limit of $2.9 \times 10^{-11} \mu_B$ at a $90 \%$ C.L~\cite{Beda:2012zz}, while the TEXONO collaboration has determined the limit $\mu_\nu < 7.4 \times 10^{-11}\mu_B$~\cite{TEXONO:2006xds}. In solar neutrino experiments, the following constraints have been reported: $\mu_\nu < 1.1\times 10^{-10} \mu_B$ from the Super-Kamiokande experiment~\cite{Super-Kamiokande:2004wqk} and $\mu_\nu < 5.4\times 10^{-11} \mu_B$ from the Borexino experiment~\cite{Borexino:2008dzn}. The XENON1T experiment reported an unexpected excess in electron recoil events~\cite{XENON:2020rca}. This anomaly suggested a possible effective neutrino magnetic moment in the range of $(1.4,2.9)\times 10^{-11}\mu_B$ as a potential explanation~\cite{Miranda:2020kwy,Babu:2020ivd}. However, after an upgrade to the detector, systematic uncertainties were significantly reduced, resulting in a decreased of the background by more than $50 \%$. With the new data collected by the XENONnT collaboration, the electronic recoil has been observed with no excess in the range $(1-7) $ keV, and the new constraint on the effective magnetic moment $\mu_\nu < 6.4 \times 10^{-12}$  at $90\%$ C.L. has been reported~\cite{XENON:2022ltv}. Simultaneously, the LUX-ZEPLIN (LZ) collaboration, which focuses on the search for dark matter candidates, has released its initial results based on an exposure of 5.5 tons over 60 live days of liquid Xenon~\cite{LZ:2022lsv}. This new data from LZ can be used to set a stringent limit on effective neutrino magnetic moment: $\mu_\nu < 6.2 \times 10^{-12} \mu_B$.\\

In this work, we employ scalar Leptoquark (LQ) interactions to generate a significant magnetic moment for Majorana neutrinos that may fall within the  reach of current experimental capabilities. The analysis of Leptoquarks is acquiring importance due to their potential to explain specific anomalies, such as the discrepancy observed in semileptonic $B$ meson decays~\cite{Mandal:2018kau, Aydemir:2019ynb}. The $R_{k^(*)}$ anomalies have also been investigated using LQ models~\cite{Becirevic:2017jtw,Becirevic:2016yqi}; however, the current measurement of the $b \to s\ell^+\ell^-$ decay, carried out by the LHCb collaboration, appears to be consistent with the SM predictions~\cite{LHCb:2022qnv}. A similar situation exists concerning the muon magnetic moment, where scalar LQs have been employed to address the discrepancy $\Delta a_\mu$ \cite{Bigaran:2020jil,Dorsner:2020aaz,Baek:2015mea,ColuccioLeskow:2016dox}. Nevertheless, recent theoretical calculations are aligning with the experimental data~\cite{Muong-2:2025xyk,Aliberti:2025beg}. In this regard, a recent analysis about the implications of the final result from the Fermilab $(g-2)_\mu$ experiment in beyond the Standard Model theories has been done in Ref.~\cite{Athron:2025ets}, where constraints on the mass and couplings of the LQ $\slq$ have been derived. Another motivation for considering LQs is their ability to generate a neutrino mass term through one-loop processes~\cite{Saad:2020ihm,Popov:2016fzr,Cai:2017wry,AristizabalSierra:2007nf}. In contrast, to achieve significant values for neutrino magnetic moments, one would need to carefully adjust the parameters to satisfy the demands of the neutrino mass pattern. Several mechanisms within LQ models have been proposed~\cite{Brdar:2020quo}, which typically involve introducing at least two LQ states in order to produce a substantial magnetic moment, while keeping the neutrino mass below the eV range. Previous studies have explored the contribution of LQs to neutrino magnetic moments. For instance, the authors of Ref.~\cite{Chua:1998yk} examined vector Leptoquarks and estimated the resulting neutrino magnetic moment to be on the order of $10^{-10}\mu_B$ ($10^{-12}\mu_B$) for third-(second-)generation LQs. Furthermore, Ref.~\cite{Povarov:2007zz} explored the phenomenology of scalar LQ within a minimal model that incorporates four-color symmetry. This study established constraints on the LQ mass, finding that $m_{LQ}>1.45$ TeV based on the astrophysical constraint on the neutrino magnetic moment $\mu_\nu < 3\times 10^{-12}\mu_B$~\cite{Raffelt:1996wa}. Furthermore, a more comprehensive investigation of electromagnetic properties of neutrinos, considering the contributions of both scalar and vector Leptoquarks, was carried out in~\cite{Bolanos-Carrera:2023ppu}. In this analysis, a neutrino magnetic moment on the order of $\mu_\nu \sim \mathcal{O}(10^{-9})\mu_B$ was derived for neutrino masses in the eV range. Also, the neutrino magnetic moment has been studied more recently in~\cite{Brdar:2020quo}, assuming the existence of right-handed neutrinos that are heavier than the left-handed SM neutrinos. The authors explore how  scalar LQs contribute to the neutrino magnetic moment, particularly within a framework that maintains an exact $SU(2)_H$ symmetry. On the other hand, experimental data significantly restrict the masses and couplings of vector LQs \cite{Valencia:1994cj,Kuznetsov:1994tt}, which is why scalar LQ analysis has been favored in the literature.\\

In this paper, we investigate the neutrino magnetic moment within a framework where the SM is augmented with the scalar LQs $S_1(\bar{3},1,1/3)$ and $\widetilde{R}_2(3,2,1/6)$, commonly referred to as the doublet-singlet scalar Leptoquark (DSL) model. To our knowledge, the neutrino magnetic moment has not been analyzed in the context of the DSL model before. This model has been considered in \cite{Zhang:2021dgl,Dev:2024tto,Parashar:2022wrd,Dorsner:2017wwn} due to its potential to generate masses for Majorana neutrinos. Additionally, they may provide insights for lepton flavor universality violation of B-meson, such as the one defined by the ratio $R_{D^{(*)}} = \text{Br}(\bar{B}\to D^{(*)}\tau\bar{\nu})/\text{Br}(\bar{B}\to D^{(*)}\ell\bar{\nu})$ with $\ell=e,\mu$. In the literature, the triplet-doublet model, which extends the SM with the LQs $S_3(\bar{3},3,1/3)$ and $\widetilde{R}_2(3,2,1/6)$, has also been studied for generating a neutrino mass term. However, it is not possible to address the $R_{D^{(*)}}$ anomaly with the $S_3$ Leptoquark, reason why we prefer the DSL model. Also, in the DSL model, the mixing between $S_1$ and $\widetilde{R}_2$, induced by a Higgs interaction, produces an enhancement in the neutrino magnetic dipole moment. This allows the neutrino magnetic moment to approach values close to current experimental values. In addition, we also examine other well-studied processes induced by LQs, such as the muon anomalous magnetic moment and the Lepton Flavor-Violating (LFV) decay $\tau \to \mu \gamma$. These processes are used to investigate the parameter space for the LQ model, which can be relevant for analyzing the neutrino magnetic moment. In this context, we will concentrate on the transition magnetic moment $\mu_{\nu_{\tau \mu}}$.\\

The organization of the paper is as follows: In Section~\ref{sec:LQmodel} we briefly discuss the framework of the LQ model that we are interested in. Section~\ref{sec:neutrinomagnetic} presents a general calculation on the Majorana neutrino magnetic moment induced by the scalar Leptoquarks $S_1-\widetilde{R}_2$. In Section~\ref{sec:constraints}, we discuss the constraints on LQ couplings based on experimental data, followed by a numerical analysis of the neutrino magnetic moment in Section~\ref{sec:Analysis}. Finally, the conclusions and perspectives are presented in Section.~\ref{sec:Summary}. 

\section{The doublet-singlet LQ model}\label{sec:LQmodel}

Leptoquarks naturally arise in the context of Grand Unified Theories (GUTs)~\cite{Pati:1974yy,Georgi:1974sy,Fritzsch:1974nn}, where strongly non-interacting leptons are accommodated into the same multiplets as quarks. Other well-established theoretical frameworks predicting the existence of LQs include technicolor models~\cite{Ellis:1980hz,Farhi:1980xs,Hill:2002ap}, $R$-parity violating supersymmetric models~\cite{Barbier:2004ez}, and models with composite fermions~\cite{Schrempp:1984nj,Buchmuller:1985nn,Gripaios:2009dq}. These theoretical particles can be either color-triplet scalars or bosons, and their main characteristic is to convert leptons into quarks and vice versa. The physics of Leptoquarks can be systematically studied  based on their representation under the SM gauge group~\cite{Buchmuller:1986zs}, where ten different LQ states emerge if the SM is permitted to have purely left-handed neutrinos. However, more LQs arise if electrically neutral states, that play the role of right-handed neutrinos, are added to the SM particle spectrum. Leptoquark phenomenology is usually explored using a model-independent approach based on an effective Lagrangian, which allows us to focus on the low-energy LQ interaction, while ignoring the complexities of ultraviolet completion. The most general Lagrangian of dimension four with effective interactions and invariant under $SU(3)_c \times SU(2)_L \times U(1)_Y$ for both scalar and vector LQs, was first presented in~\cite{Buchmuller:1986zs}. For a more recent review, we recommend consulting the Ref.	~\cite{Dorsner:2016wpm}. We focus on a model with two scalar Leptoquarks: a singlet LQ, denoted as $S_1(\bar{3}, 1, 1/3)$, and a doublet LQ under $SU(2)$, denoted as $\widetilde{R}_2(3,2,1/6)$. The effective Lagrangian associated with the LQs $S_1$ and $\widetilde{R}_2$ is given by
\begin{equation}
\begin{split}
\mathcal{L}_{LQ}&= \left( y_{i \alpha}^L \overline{Q}^{ \,c}_{i L} \epsilon \ell_{\alpha L} S_1+y_{i \alpha}^R \overline{u}_{i R}^c e_{\alpha R} S_1+y_{i \alpha} \overline{d}_{i R} \widetilde{R}_2^T \epsilon \ell_{\alpha L}+\text { h.c. }\right) \\
& +\left(D_\mu S_1\right)^{\dagger}\left(D^\mu S_1\right)+\left(D_\mu \widetilde{R}_2\right)^{\dagger}\left(D^\mu \widetilde{R}_2\right)-V_{LQ},
\end{split}
\end{equation}
where $\overline{Q}^{\,c}_{i L}$ and $\ell_{\alpha L}$ denote the left-handed quark and lepton doublets, respectively, with flavor indices $i$ and $\alpha$. Besides, $\overline{u}^{\,c}_{iR}$ ($\overline{d}_{iR}$) and $e_{\alpha R}$ are the right-handed up-type (down-type) quark and charged lepton singlets, respectively. The superscript $c$ in the fermion fields stands for the charge conjugation field, defined as 
\begin{equation}\label{eq:CCFields}
\Psi^c= C\bar \Psi^T, \quad \text{and} \quad \bar \Psi^c=-\Psi^T C^{-1},
\end{equation}
with $C$ the charge conjugation matrix. The Yukawa couplings $y_{i\alpha}^{L,R}$ and $y_{i\alpha}$ represent the LQ coupling with a quark from generation $i$ and a lepton from generation $ \alpha$. The  scalar potential for the LQs $S_1$ and $\widetilde{R}_2$, neglecting the quartic terms since they are irrelevant for our calculation, is given by
\begin{equation}\label{LQpotential}
\begin{split}
V_{LQ} &= m_1^2 S_1^{\dagger} S_1+m_2^2 \widetilde{R}_2^{\dagger} \widetilde{R}_2+\alpha_1\left(H^{\dagger} H\right)\left(S_1^{\dagger} S_1\right)+\alpha_2\left(H^{\dagger} H\right)\left(\widetilde{R}_2^{\dagger} \widetilde{R}_2\right)\\
&+\alpha_2^{\prime}\left(H^{\dagger} \widetilde{R}_2\right)\left(\widetilde{R}_2^{\dagger} H\right)  +\left(\kappa H^{\dagger} \widetilde{R}_2 S_1+\text { h.c. }\right),
\end{split}
\end{equation}
where the coefficients $\alpha_{1,2}$ and $\alpha^\prime$ are real couplings that describe the strength of quartic interactions between the LQs and the SM Higgs doublet. After electroweak symmetry breaking, the trilinear coupling $\kappa$ leads to a mixing between $S_1$ and the LQ doublet state with electric charge 1/3. This mixing is necessary to achieve lepton number violation and to generate a neutrino mass term radiatively. To avoid a proton rapid decay, one can assign $B = -1/3$ to $S_1$ and $B = 1/3$  to $\widetilde{R}_2$ to ensure the absence of $B$-violating terms in the Lagrangian. The LQ mass matrix is 
\begin{equation}
M^2_\text{mix} = \left( \begin{array}{r r}
						m_{S}^2 & \frac{v}{\sqrt{2}}\kappa \\
						\frac{v}{\sqrt{2}}\kappa & m^2_{R}
						\end{array} \right),
\end{equation}
where $m^2_{S} = m_1 ^2+\alpha_1 v^2/2$ and $m^2_{R} = m_2^2 +(\alpha_2+ \alpha_2^\prime)v^2/2$, with $v$ the vacuum expectation value of the Higgs boson. The LQ mass matrix can be diagonalized by a rotational matrix, which is parametrized by the mixing angle $\theta_{LQ}$ and get the physical mass eigenstates	
\begin{equation}
\begin{split}
\slq & = \cos \theta_{LQ} S_1 - \sin\theta_{LQ} \widetilde{R}_2^{-\frac{1}{3}*} ,\\
\rlq & = \sin \theta_{LQ} S_1 + \cos\theta_{LQ} \widetilde{R}_2^{-\frac{1}{3}*}. \\
\end{split}
\end{equation}
where the mixing angle is given in terms of the mass eigenstates as $\tan 2\theta_{LQ}= \sqrt{2}\kappa v /(m_{R}^2 - m^2_{S})$. The corresponding mass eigenvalues are 
\begin{equation} 
m^2_{\slq,\rlq} = \frac{1}{2}\left(m^2_{S} + m ^2_{R}\mp \sqrt{\left( m^2_{S}-m ^2_{R} \right)^2 + 2\kappa^2 v^2} \right).
\end{equation}

For the Leptoquark with electric charge 2/3, the corresponding mass term is given by 
\begin{equation}
m^2_{R^{2/3}} = m^2_2+\frac{1}{2}\alpha_2 v^2.
\end{equation}
Rotating the Lagrangian from the weak to the mass basis for quarks and leptons, the interaction terms take the form 
\begin{equation}\label{eq:LagS1}
\begin{aligned} 
\mathcal{L}_{\mathrm{Y}} &=  \overline{\nu}_\alpha\left(y_{i \alpha}^* \sin \theta_{LQ} P_{R}-y_{i \alpha}^{L} \cos \theta_{LQ} P_{L}\right) d_i \slq\\
&+\overline{l_\alpha^c}\left(y_{i \alpha}^{\prime L} P_{L}+y_{i \alpha}^{R} P_{R}\right) \cos \theta_{LQ} u_i \slq \\
&-\overline{\nu}_\alpha\left(y_{i \alpha}^* \cos_{LQ} \theta P_{R}+y_{i \alpha}^{L} \sin \theta_{LQ} P_{L}\right) d_i \rlq\\
&+\overline{l_\alpha}\left(y_{i \alpha}^{L} P_{L}+y_{i \alpha}^{R} P_{R}\right) \sin \theta_{LQ} u_i \rlq + y_{i \alpha} \bar{d}_i P_{L} l_\alpha R^{2/3}+\text { h.c. }
\end{aligned}
\end{equation}
with $y^\prime  = V^Ty^L$, since we choose the down-type quark basis, where the left-handed quark doublet is $Q_i = ( \begin{matrix} (V^\dagger u_L)_i & d_{i L} \end{matrix} )$. 
The complete Lagrangian, including all the LQs representations and the corresponding Feynman rules, can be found in Ref.~\cite{Crivellin:2021ejk}, which can be used for an automated analysis of Leptoquarks. 

\section{Leptoquark contribution to the neutrino magnetic moment and neutrino masses}
\label{sec:neutrinomagnetic}

\begin{figure}[h]
\centering
\includegraphics[width=16cm]{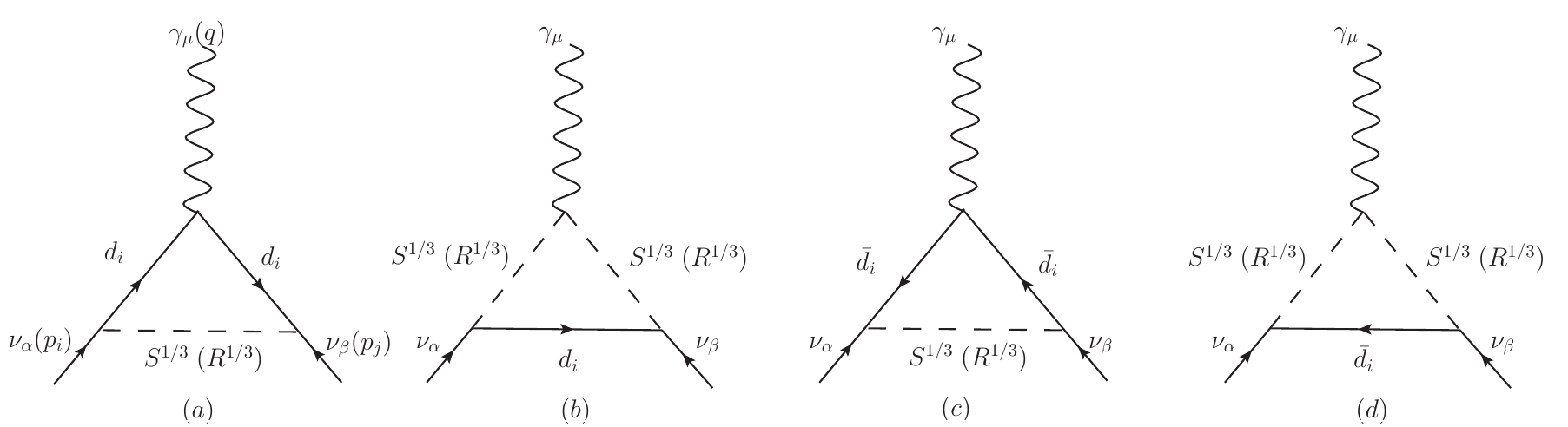}
\caption{One loop diagrams representing the scalar Leptoquarks $\slq$ and $\rlq$ contribution to the transition neutrino magnetic moment. The arrows indicate the fermion flow and the convention for the four-momenta is depicted in the diagram (a).}\label{fig:FeynmanDiag1} 
\end{figure}

It has been suggested that a neutrino magnetic moment of the order of $\mathcal{O}(10^{-12})\mu_B$ is favored for Majorana neutrinos; therefore, in this work, we focus on the transition magnetic moment $\mu^M_{\nu_{\alpha\beta}}$. In the DSL model, the Majorana neutrino magnetic moment is derived from the Feynman diagrams shown in Fig.~\ref{fig:FeynmanDiag1}, where $\nu_\alpha,\nu_\beta=\nu_e,\nu_\mu,\nu_\tau$.
 
As illustrated in Fig.~\ref{fig:FeynmanDiag1}, two fermion flows converge at a vertex, which requires a special approach. For interactions involving charge-conjugate SM fermions, we utilize the methodology described in Ref.~\cite{Crivellin:2021ejk}. Afterwards, we apply the Feynman parametrization technique to evaluate the corresponding amplitudes, enabling us to express the generalized form of a neutrino electromagnetic vertex function
\begin{align}
\mathcal M^\mu_{\alpha \beta} &= \bar u(p_j) \Bigl[ \left(\gamma^\mu - q^\mu \slashed{q} /q^2\right) \left(f^1_{\alpha \beta }(q^2) +f^2_{\alpha \beta} (q^2) q^2 \gamma^5 \right)\nonumber\\
& -i \sigma^{\mu \nu} q_\nu \left(f^3_{\alpha \beta} (q^2) + i f^4_{\alpha \beta} (q^2) \gamma^5\right) \Bigr] u(p_i).
\end{align}
The neutrino magnetic moment form factor is defined as $\mu_{\nu_{\alpha \beta}} = f^3_{\alpha \beta} (q^2)$ when coupled with a real photon at $q^2=0$ (static magnetic moment).  The first two diagrams contribute to the magnetic moment of the Dirac neutrino, which can be expressed as
\begin{equation}\label{eq:neutrinomoment_Dirac}
\mu^D_{\nu_\alpha \beta}=\mu^{D,\slq}_{\nu_\alpha \beta}+\mu^{D,\rlq}_{\nu_\alpha \beta}
\end{equation}
with
\begin{equation}
\begin{split}
\mu^{D,\slq}_{\nu_{\alpha \beta}} &=-\frac{N_c m_e \mu_\text{B}}{16 \pi^2 m^2_{\slq}}\sum^3_{i=1} \Big[- 2m_{d_i}\sin(2\theta_{LQ}) \left( y^{L*}_{i\alpha}y^*_{i\beta}+y^L_{i\beta}y_{i\alpha} \right)\mathcal{G}\Big( \frac{m^2_{d_i}}{m^2_{\slq}}\Big) \Big ],\\
\mu^{D,\rlq}_{\nu_{\alpha \beta}} &=-\frac{N_c m_e \mu_\text{B}}{16 \pi^2 m^2_{\rlq}}\sum^3_{i=1} \Big[2m_{d_i}\sin(2\theta_{LQ}) \left(y^{L*}_{i\alpha}y^*_{i\beta}+y^L_{i\beta}y_{i\alpha} \right)\mathcal{G}\Big(\frac{m^2_{d_i}}{m^2_{\rlq}}\Big) \Big].
\end{split}
\end{equation}
where $m_{d_i}=(m_d,m_s,m_b)$ is the down-type quark mass and the function $\mathcal{G}$ is given by 
\begin{align}
\mathcal{G}(a) & = \frac{a-1-\ln(a)}{6 (a - 1)^2}.
\end{align}

In the Majorana case the diagrams c) and d) also contribute to the magnetic moment~\cite{Shrock:1982sc}. Therefore, the Majorana neutrino magnetic moment induced by the LQs $\rlq$ and $\slq$ can be expressed as follows:
\begin{equation}
\mu^M_{\nu_\alpha \beta}=\mu^{M,\slq}_{\nu_\alpha \beta}+\mu^{M,\rlq}_{\nu_\alpha \beta}
\end{equation}
with
\begin{equation}\label{eq:neutrinomoment}
\begin{split}
\mu^{M,\slq}_{\nu_{\alpha \beta}} &=-\frac{i N_c m_e \sin(2\theta_{LQ}) \mu_\text{B}}{8 \pi^2 m^2_{\slq}}\sum^3_{i=1} \Big[2m_{d_i} \text{Im} \left( y^{L*}_{i\beta}y^*_{i\alpha}-y^{L^*}_{i\alpha}y^*_{i\beta} \right)\mathcal{G}\Big( \frac{m^2_{d_i}}{m^2_{\slq}}\Big)\Big ],\\
\mu^{M,\rlq}_{\nu_{\alpha \beta}} &=\frac{i N_c m_e \sin(2\theta_{LQ})\mu_\text{B}}{8 \pi^2 m^2_{\rlq}}\sum^3_{i=1} \Big[2m_{d_i} \text{Im} \left(y^{L*}_{i\beta }y^*_{i\alpha }-y^{L*}_{i\alpha}y^*_{i\beta} \right)\mathcal{G}\Big(\frac{m^2_{d_i}}{m^2_{\rlq}}\Big) \Big].
\end{split}
\end{equation} 

As expected, $\mu^M_{\nu_{\alpha\beta}}$ is both antisymmetric and imaginary. Generally, without introducing right-handed neutrinos into the particle spectrum of the DSL model, the neutrino magnetic moment is enhanced by the mixing of the scalar LQs, since a proportional term to the quark mass running inside the loop is obtained. In contrast, other LQ models typically require the addition of right-handed neutrinos to achieve a significant neutrino magnetic moment~\cite{Bolanos-Carrera:2023ppu,Chua:1998yk,Povarov:2007zz}. For instance, in~\cite{Bolanos-Carrera:2023ppu}, a diagonal neutrino magnetic moment for Dirac neutrinos was derived, showing dependence on both the left and right LQ-Yukawa couplings, similar to Eq.~\eqref{eq:neutrinomoment_Dirac}.\\

In the DSL model, neutrinos are massless at the tree level. However, due to the trilinear coupling $\kappa$ in Eq.~\eqref{LQpotential}, the LQs $\slq$ and $\rlq$ mix, and this produces a non-trivial lepton number violating Majorana mass matrix for neutrinos through loop corrections. The Feynman diagrams that contribute to the neutrino mass can be obtained by removing the photon line from the diagrams shown in Fig.~\ref{fig:FeynmanDiag1}. Consequently, the Majorana neutrino mass, induced by the LQs $\slq$ and $\rlq$, can be expressed as~\cite{Mahanta:1999xd,Zhang:2021dgl}
\begin{align}\label{eq:neutrinomass}
m_{\nu_{\alpha\beta}}&=\frac{N_c}{32\pi^2}\sin(2\theta_{LQ})\sum_{i=1}^3 m_{d_i}(y^{L*}_{i\alpha}y^*_{i\beta}+y^{L*}_{i\beta}y^*_{i\alpha}) \int^1_0 d\,x \Big( \ln (m^2_{d_i}x-m^2_{\rlq}(x-1))\nonumber \\
&- \ln (m^2_{d_i}x-m^2_{\slq}(x-1)) \Big) ,\nonumber \\
&=\frac{N_c}{32\pi^2}\sin(2\theta_{LQ})\sum_{i=1}^3 m_{d_i}(y^{L*}_{i\alpha}y^*_{i\beta}+y^{L*}_{i\beta}y^*_{i\alpha})\ln\left(\frac{m^2_{\rlq}}{m^2_{\slq}}\right),
\end{align}
where as usual, the limit $m_{d_i}/m_{LQ}\to 0$ has been considered. Then, the mixing between the LQs is important for generating neutrino mass of Majorana type, as well as generating magnetic moments for Majorana neutrinos. Also, as noted in the neutrino mass matrix, if the LQs masses are exactly degenerate, their contribution to neutrino mass will cancel.

\section{ Constraints on the parameter space of the scalar LQ model}
\label{sec:constraints}
In this section, we provide the treatment of the parameter space for the LQ model described in Section~\ref {sec:LQmodel}. We first discuss the latest limits on the LQ mass imposed by ATLAS and CMS experiments. After that, we examine various processes to restrict the LQ couplings to fermions. Specifically, we consider the muon magnetic moment, and the experimental limit on the LFV decay Br($\tau \to \mu \gamma$).
\subsection{Constraints on the LQ mass}

Although all particles predicted by the SM have been experimentally detected, extensive efforts have been made to uncover additional signals that point out the path towards a complete theory of the fundamental interactions. Namely, the search for Leptoquarks has been carried out in numerous experiments, and to date, no signals have been detected. However, limits on their properties, such as the LQ mass and its couplings to fermions, can be imposed by the data. Typically, one can employ the theoretical prediction of the LQ cross section to derive experimental upper limits, which can be interpreted as lower limits on the LQ mass. The search for a scalar LQ with an electric charge of $1/3\, e$ is driven by models that can explain various anomalies in $B$ meson decays, where the LQ interacts with third-generation fermions. ATLAS and CMS collaborations have investigated LQs searches in proton-proton collisions at $\sqrt{s} = 13$ TeV, focusing on both pair and singly production mechanisms. Assuming that LQs are pair-produced and can only decay into $t \tau $ and $b \nu$ channels, the ATLAS collaboration as set a constraint on the LQ mass $m_{LQ} > 1000$ GeV based on data from the second LHC run with an integrated luminosity of  $36.1 \; \text{fb}^{-1}$~\cite{ATLAS:2019qpq}. The CMS collaboration ruled out masses below $900$ GeV at a $95 \%$ confidence level, considering pair production of LQs that exclusively couple to third-generation fermions, specifically with Br$(LQ \to t \tau)=1$~\cite{CMS:2018svy}. For the decay channel $LQ \to b \nu$, a mass range of $m_{LQ} < 1100$ GeV is excluded, while for $t \nu$ channel, the mass satisfies $m_{LQ} > 1020$~\cite{Takahashi:2018qwa}. Second-generation LQs have also been explored, where events are selected by detecting a pair of oppositely charged muons and at least two jets produced by charm or bottom quarks. Assuming Br$(LQ \to c \mu)=1$, ATLAS has set the constraint of $m_{LQ}>1700$ GeV at $95 \%$ C.L. under the scenario that the LQ is pair-produced~\cite{ATLAS:2020dsk}. The CMS detector has also investigated the constraints on second-generation Leptoquarks, where LQs with masses below 1530 and 1285 GeV have been excluded~\cite{CMS:2018lab}. Other experimental constraints on the masses and couplings of second-generation LQs have been established through direct searches at the CERN LHC, operating at a collision energy of $\sqrt{s} = 13$ TeV~\cite{CMS:2018oaj,ATLAS:2020xov,CMS:2022nty,ATLAS:2022wcu}. Recent studies have approached the problem by simultaneously considering both pair and single LQ production mechanisms, represented as $\sigma(pp \to  LQ \overline{LQ}) + \sigma(pp \to \ell LQ)$,  where the decays $LQ \to (t \tau, b\nu)$ are allowed. In this context, the CMS experiment has found a lower limit on the LQ mass, ranging from $980$ to $1730$ GeV in proton-proton collisions with a center-of-mass energy of $\sqrt{s}=13$ TeV and an integrated luminosity of $137 \; \text{fb}^{-1}$~\cite{CMS:2020wzx}. For an easy reference, we present in Tab~\ref{tab:massbounds} the most recent constraints on the mass of a scalar Leptoquark with electric charge $1/3\;e$.
 
\begin{table}[t]
\begin{center}
\begin{tabular}{c  c  c c}
\hline  \hline
Experiment	& Channel &Lower bound (95$\%$ C.L) & Ref  \\
\hline \hline
ATLAS &$LQ\to \mu q$       &$1.8-2.0 $ TeV &\cite{ATLAS:2022wcu}\\
ATLAS &$LQ \to \tau b / t\tau$       &$1.0-1.4$ TeV  &\cite{ATLAS:2019qpq}\\
CMS   &$LQ\to  t\tau / b\nu$   &$0.98-1.73$ TeV  &\cite{CMS:2020wzx}\\
CMS   &$LQ \to t\tau /t\mu$   &$1.12-1.42$ TeV  &\cite{CMS:2022nty} \\
\hline \hline
\end{tabular}
\caption{Lower bounds on the mass of a LQ with electric charge of $1/3 \; e$, imposed by the CMS and ATLAS collaborations.  For details about the generation of LQs and assumption regarding their decays modes, we suggest to consulting the corresponding references.}
\label{tab:massbounds}
\end{center}
\end{table}

The coupling of Leptoquarks to fermions is also constrained by Drell-Yan processes $pp \to \ell \ell$ and $pp \to \ell \nu$, as demonstrated in~\cite{Allwicher:2022gkm}. The analysis shows that the restriction $\sqrt{y^R_{22}y_{22}}<0.66$ has been established using the latest LHC data. On the other hand, we also consider the bounds $y^L_{3\alpha}, y^R_{3\alpha} \leq 1$ in our analysis, which are imposed by the LHC results, and to avoid the breakdown of perturbation theory~\cite{Bolanos-Carrera:2023ppu}. Due to experimental restrictions previously mentioned, in the following analysis, we consider two values of the  LQ mass: $m_{LQ}=1.5$ TeV and $m_{LQ}=2$ TeV. For the remaining parameters, we set $m_{1} = m_{2}=m_{LQ}$, $\alpha_1 = \alpha_2 = \alpha^\prime_2 = 0.2$ and $\kappa = 50$ GeV~\cite{Parashar:2022wrd}. These values produce the following results:
\begin{equation}
\begin{split}
m_{LQ} &= 1500 \; \text{GeV}:\; m_{\slq}=1499\;\text{GeV},\; m_{\rlq}=1506\;\text{GeV},\; \theta_{\text{LQ}} = 0.617\;\text{rad}\\
m_{LQ} &= 2000 \;\text{GeV}:\; m_{\slq}=1999\;\text{GeV},\; m_{\rlq}=2004\;\text{GeV},\; \theta_{\text{LQ}} = 0.617\;\text{rad}
\end{split}
\end{equation}

\subsection{Magnetic moment of the muon and LFV $\tau \to\mu \gamma$ constraints}
\label{subsec:space-parameter}
Scalar Leptoquarks can significantly affect certain observables that have been accurately measured. Among the most critical processes are the muon magnetic moment and the LFV decays $\ell_i \to \ell_j \gamma$. Recently, the muon anomalous magnetic moment was updated by the Fermi National Accelerator Laboratory (FNAL), reporting a value of $a_\mu =1165920710(162)\times 10^{-12}$(139 ppb)~\cite{Muong-2:2025xyk}, which combined with previous results, the world average is $a_\mu(\text{exp})=1165920715(145)\times 10^{-12}$(124 ppb). On the theoretical side, there is also an important update; new progress in calculating the hadronic light by light scattering contribution provides the standard model value $a^{SM}_\mu=116592033(62)\times 10^{-11}$(530 ppb)~\cite{Aliberti:2025beg}. With these new experimental and theoretical data, the difference is $a^{exp}_\mu-a^{SM}_\mu= 38(63)\times 10^{-11}$, indicating that there is no longer tension between the SM and the experimental value. With the disappearance discrepancy $\Delta a_\mu$, the process the muon anomalous magnetic moment is now regarded as a high-precision tool for constraining the parameter space of BSM theories, particularly for the DSL model. In the following  analysis of the parameter space, we use the most up-to-date values of $(g-2)_\mu$. 

On the other hand, ongoing experimental investigation of LFV decays has established strict limits on their branching ratios. These limits can place constraints on the parameters that extend the Standard Model. The BaBar collaboration reported an upper limit of Br$(\tau \to \mu \gamma) < 4.4\times 10^{-8}$ at $90\%$ C.L.~\cite{BaBar:2009hkt}. Although the LQ contribution to both processes has been extensively studied in the literature~\cite{Cheung:2001ip,Dorsner:2020aaz,Cheung:2015yga}, we reproduce the relevant calculations and leave the respective formulas in Appendix~\ref{appendixLVFdecay}. Since we are interested in the allowed values for the LQ Yukawa coupling to fermions, we utilize the latest measurement of the muon anomalous magnetic moment to determine the sensitivities of $y^{L,R}_{22}$ and $y^{L,R}_{32}$, while the LFV $\tau \to \mu \gamma$ decay can delimit the couplings $y^{L,R}_{23}$ and $y^{L,R}_{33}$ as well.

\begin{figure}[h]
\centering
\includegraphics[width=16 cm]{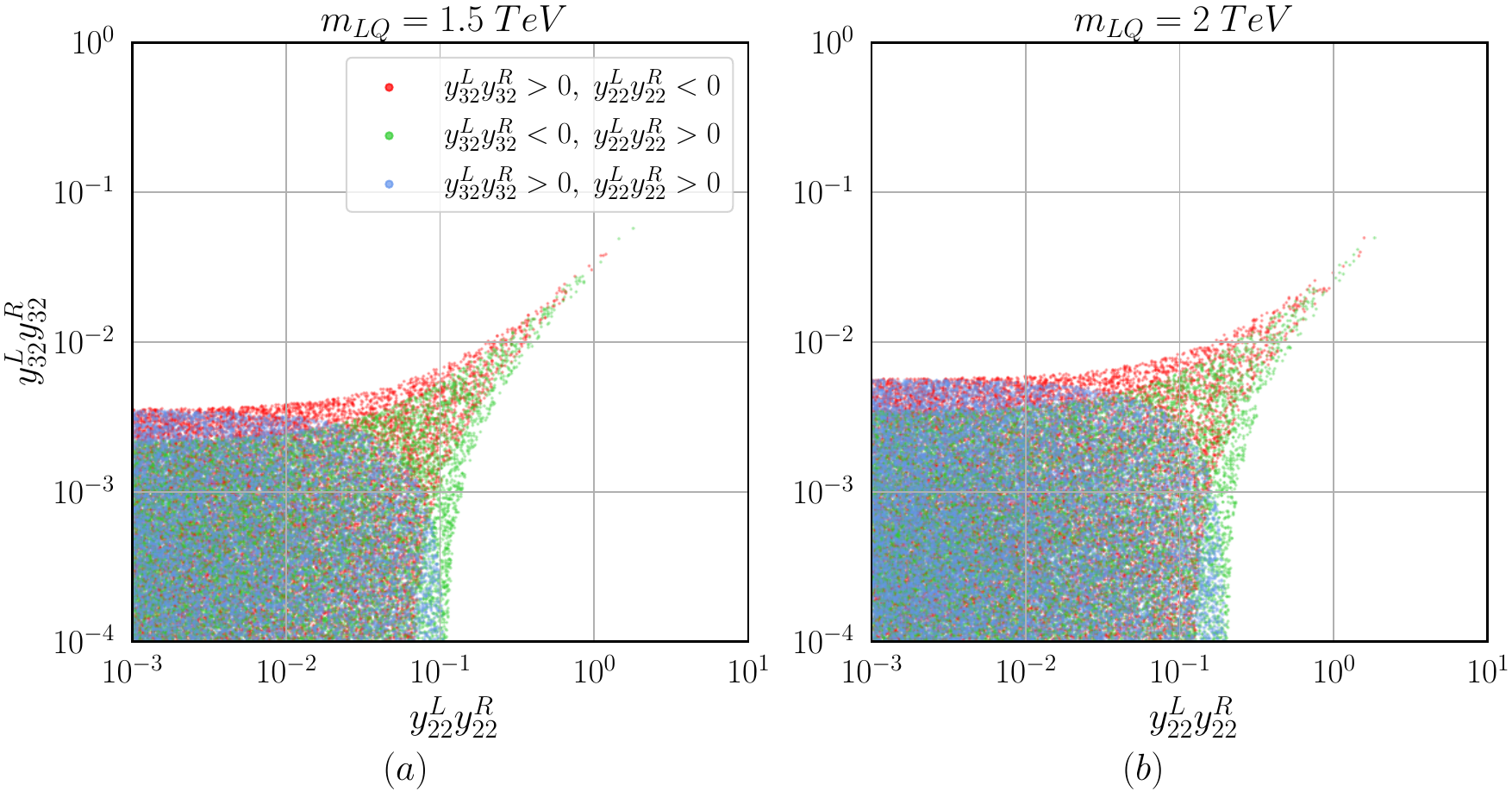}
\caption{Allowed points in the plane $y^L_{22}y^{R}_{22}$ vs $y^L_{32}y^{R}_{32}$ at $95\%$ C.L. consistent with the $(g-2)_\mu$ processes for $m_{LQ} = 1.5$ TeV (plot (a)), and $2$ TeV (plot (b)). Notice that in cases with opposite sign, a degeneracy in the parameters can lead to large values of the coupling products.}\label{fig:space_muon} 
\end{figure}

We perform a randomly scan for a few thousands of 4-tuples $\{ y^{L}_{22},y^{R}_{22},y^{L}_{32},y^{R}_{32}\}$ consistent with the recent values of $(g-2)_\mu$ at $95\%$ C.L. The allowed points are illustrated in Fig~\ref{fig:space_muon} for three different scenarios based on the relative sign of the coupling products $y^L_{32}y^{R}_ {32}$ and $y^L_{22}y^{R}_{22}$. We note that while the previous discrepancy $\Delta a_\mu$ has been resolved (indicating alignment between experimental and theoretical results), there remains a region where the LQ coupling products can be of the order of $\mathcal{O}(1)$. These large values are most favorable in scenarios where the coupling products have opposite signs, allowing for partial contributions to cancel each other out and thus facilitating larger permissible values. Similar regions where cancellations occurs have been previously noted in the literature \cite{AristizabalSierra:2021fuc} for Majorana transition moments, where the characteristics of such blind spots were thoroughly discussed.

\begin{figure}[h] 
\centering
\includegraphics[width=15 cm]{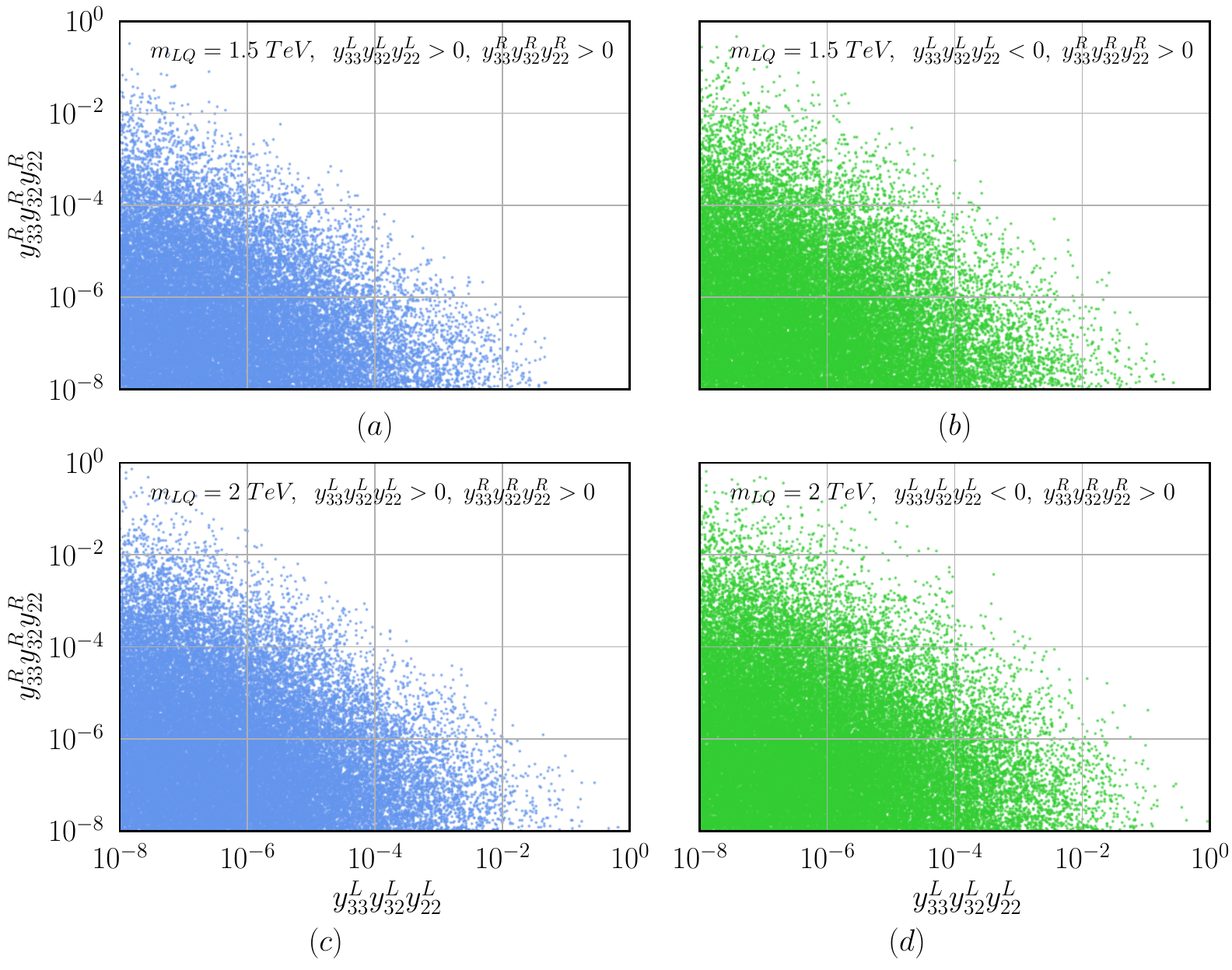}
\caption{Allowed areas on the plane $y^L_{33}y^L_{32}y^L_{22}$ vs $y^R_{33}y^R_{32}y^R_{22}$ consistent  with the $(g-2)$ muon anomaly at $95\%$ C.L. and the LFV decay $\tau\to\mu\gamma$ for a LQ mass of $1.5$ TeV (plots (a) and (b)) and $2$ TeV (plots (c) and (d)).}\label{fig:space_tau}
\end{figure}

On the other hand, the LFV decay $\tau \to \mu \gamma$ imposes an extra constraint on the couplings $y^{L,R}_{i3}$ and $y^{L,R}_{i2}$, where $i$ represents the quark generation. This constraint needs to be considered alongside the requirement stemming from the muon anomalous magnetic moment. We perform a random scan for the set of points $\{y^L_{i3}, y^R_{i3},y^L_{i2},y^R_{i2}\}$ consistent with the recent data on $(g-2)_\mu$ and experimental upper bound on the decay $\tau\to \mu\gamma$. In Fig.~\ref{fig:space_tau} we display the allowed parameter space for the triple LQ coupling products $y^L_{33}y^{L}_{32}y^{L}_{22}$ and $y^R_{33}y^{R}_{32}y^{R}_{22}$ in scenarios analogous to the muon magnetic moment analysis. Furthermore, we have considered that only the real parts of the coupling constants are non-zero. The scenario where $y^L_{33}y^{L}_{32}y^{L}_{22}<0$ and $y^R_{33}y^{R}_{32}y^{R}_{22}>0$ is slightly less constrained than the scenario where all the LQ couplings are positive. The top (bottom) panels depict results for a LQ mass of 1.5 TeV (2 TeV). Generally, as the LQ mass increases to 2 TeV, the allowed values can be slightly relaxed. This behavior is expected because the loop functions in Eqs.~\eqref{eq:muAMM} and \eqref{eq:Taudecay} are suppressed as soon as the LQ mass increases, so large values for the Yukawa couplings are needed to explain the experimental constraints. It is also important to note that various studies have explored how LQ particles can explain the discrepancies found by the BaBar and Belle collaborations in the ratios of semileptonic B meson decays $R_{D^{(*)}}$~\cite{Aydemir:2019ynb,Saad:2020ucl,Popov:2019tyc}. These processes can indeed be used to constrain the LQ parameter space; however, we have confirmed that the values of the neutrino magnetic moment $\mu_{\nu_{\mu\tau}}$ induced by the LQs remain unaffected by such restrictions

\section{Neutrino magnetic moment analysis}\label{sec:Analysis}
\begin{figure}[h] 
\centering
\includegraphics[width=17cm]{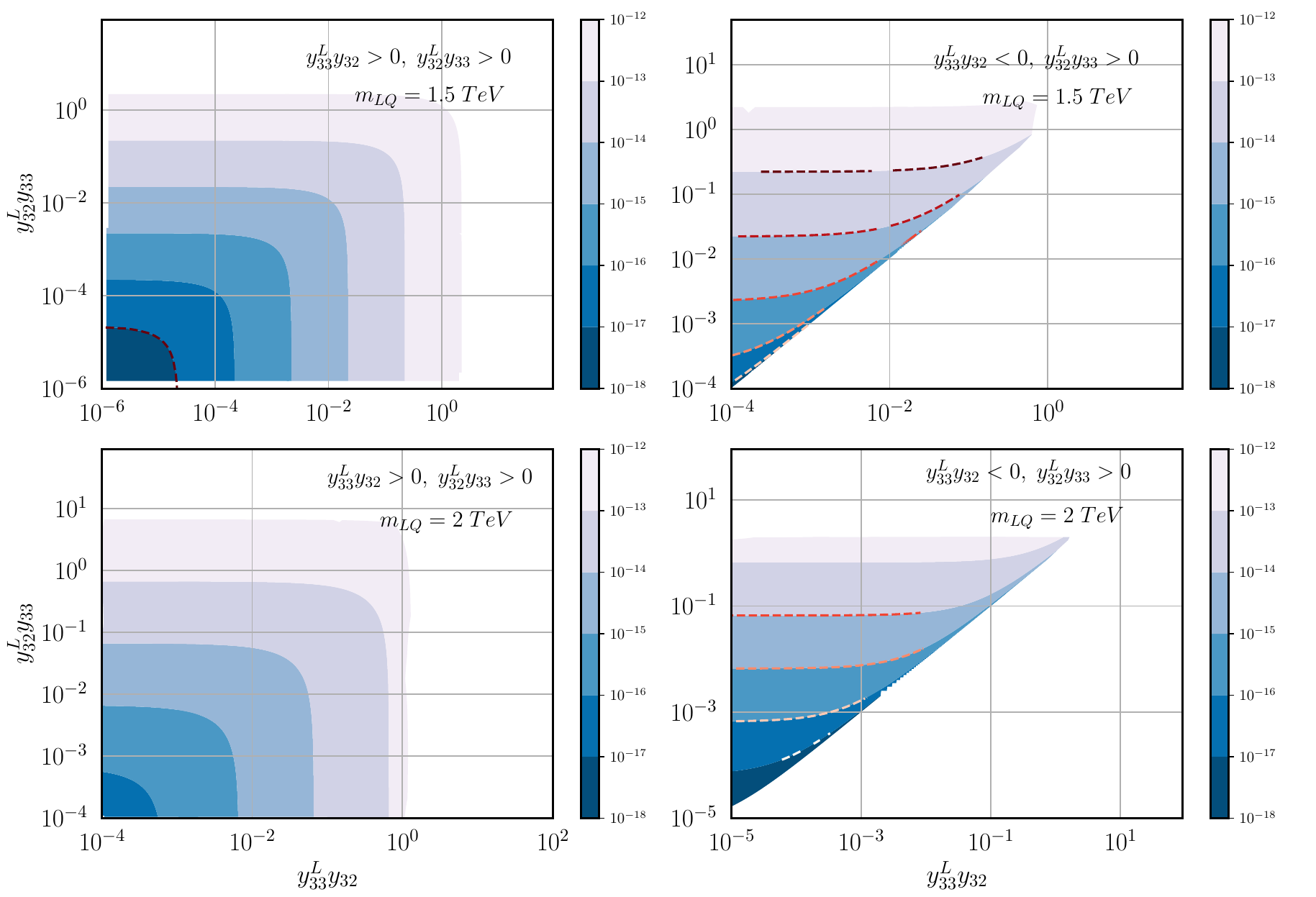}
\caption{Contours for the transition magnetic moment $|\mu^M_{\nu_{\mu\tau}}|$ in the allowed values derived from the parameter space analysis of the DSL model assuming a LQ mass of $m_{LQ}=1.5$ TeV and $2$ TeV. The dashed lines in the plots account the neutrino mass limit $\sum m_\nu \leq 0.26$ eV.}\label{fig:magnetic-neutrino} 
\end{figure}

It is evident from Eq.~\eqref{eq:neutrinomoment} that the DSL model predicts a neutrino magnetic moment that includes a term proportional to the quark mass running along the loop. This significantly increases the value of the transition magnetic moment $\mu^M_{\nu_{\alpha \beta}}$. We also observe that the neutrino magnetic moment depends explicitly on the CP phases of the Yukawa couplings through the combination of the form $\text{Im} ( y^{L*}_{i\beta}y^*_{i\alpha}-y^{L^*}_{i\alpha}y^*_{i\beta} )$. For computation simplicity, we consider a scenario in which the Yukawa couplings for the LQ $S_1$ are taken to be real, while the non-zero couplings of $\widetilde{R}_2$ are purely imaginary. Taking in to account the allowed parameter space for the LQ Yukawa couplings, we present in Fig.~\ref{fig:magnetic-neutrino} the contour plots for $|\mu^M_{\nu_{\mu\tau}}|$ for two different LQ mass values: $m_{LQ}=1.5$ and $2$ TeV. Our results explore two scenarios: one in which the products of the LQ couplings $y^L_{33}y_{32}$ and $y^L_{32}y_{33}$ have the same sign, and another in which they have opposite signs. Since the DSL model generates masses for Majorana neutrinos at one-loop level, as depicted in Eq.~\eqref{eq:neutrinomass}, we utilize the upper bound $\sum m_\nu\leq 0.26$ eV, which corresponds to neutrino mass models consistent with oscillation experiments~\cite{Loureiro:2018pdz}. The contours that respect the neutrino mass constraint are indicated by the dashed lines in the plots of Fig.~\ref{fig:magnetic-neutrino}. As observed, in the scenario $y^L_{33}y_{32}>0, \;\; y^L_{32}y_{33}>0$, the neutrino mass limit constraints the magnetic moment to values up to $\mathcal{O}(10^{-17}) \mu_B$. This result is expected since the loop diagrams generating the transition magnetic moment are closely related to the ones that induce masses for the Majorana neutrinos. Then, in the scenario with positive LQ couplings, no cancellation between the LQ couplings can occur to achieve greater values for the magnetic moment while the neutrino mass scale remains close to the eV range. On the other hand, in the scenario with  opposite sign in the LQ coupling product, such cancellation can occur, which yields to regions in the LQ parameter space where the transition magnetic moment can be of the order of $\mathcal{O}(10^{-13})\; (\mathcal{O}(10^{-14}))$ for $m_{LQ}=1.5\; (2)$ TeV. This behavior can be expected when comparing the neutrino mass term in Eq.~\eqref{eq:neutrinomass}, where the neutrino mass is proportional to a sum of the LQs coupling products. In contrast, the neutrino magnetic moment in Eq~\eqref{eq:neutrinomoment} is proportional to $\text{Im} ( y^{L*}_{i\beta}y^*_{i\alpha}-y^{L^*}_{i\alpha}y^*_{i\beta})$. Consequently, this scenario requires fine-tuning of the LQs Yukawa couplings to achieve a magnetic moment on the order of $\mu_{\nu_{\mu\tau}}\sim \mathcal{O}(10^{-13})$, while keeping the neutrino mass in the eV range.\\

On the experimental side, the XENON collaboration has reported results on electron recoil events at low energies following a total exposure of 1.16 ton$\cdot$yr. With the experimental results, the following constraints for the transition neutrino magnetic moment have been reported~\cite{Singirala:2023utf}:
\begin{equation}\label{eq:NeutrinoConstraints}
\mu_{\nu_{\mu\tau}} < 9.04 \times 10^{-12}\mu_B.
\end{equation}
Additionally, the LUX-ZEPLIN collaboration has released its initial results from the search for Weakly Interacting Massive Particles (WIMPs), utilizing an exposure of 5.5 tons over 60 live days. However, the analysis has only focused on the diagonal neutrino magnetic moment, yielding the constraint $\mu_{\nu_{eff}}<1.1\times 10^{-11}\mu_B$~\cite{AtzoriCorona:2022jeb}. On the other hand, the sensitivity to electromagnetic neutrino properties for the upcoming Darwin experiment has been analyzed in~\cite{Giunti:2023yha}, where the restriction $\mu_{\nu_{eff}}<4\times10^{-12}\mu_B$ has been derived by assuming an exposure of 30 ton-years. It is evident from our analysis that the values of the neutrino magnetic moment are on the order of $10^{-13}\mu_B$ when the neutrino mass bound is taken into account, which falls below the current experimental limits.
\section{Summary and outlook}\label{sec:Summary}

A general expression for the transition magnetic moment for Majorana neutrinos has been derived in a model where the SM is extended with two colored charged scalars Leptoquarks $S_1(\bar{3},1,1/3)$ and $\widetilde{R}_2(3,2,1/6)$. After the electroweak symmetry breaking, the interaction between the LQs and the Higgs boson leads to a mixing among the Leptoquark states. Consequently, the LQs with an electric charge of $1/3 e$ produce a significant chiral enhancement in the neutrino magnetic moment, particularly due to the LQ-bottom quark contribution in the loop. Given that the lepton flavor violating decay $\mu\to e\gamma$ imposes stringent constraints on the LQ couplings to first-generation fermions, we focused on the neutrino transition magnetic moment $\mu^M_{\nu_{\mu\tau}}$. For the parameter space analysis, we consider two LQ mass values: $1.5$ and $2$ TeV, both of which are consistent with LQ searchers at the LHC through pair and single LQ production. Next, we evaluate the transition magnetic within the regions allowed by the most recent measurement of $(g-2)_\mu$, as well as constraints from $\tau \to \mu\gamma$. The evaluation of $\mu^M_{\nu_{\mu \tau}}$ was carried out under two scenarios based on the relative signs of the Yukawa couplings $y^L_{32} y_{33}$ and $y^L_{33} y_{32}$. In the favored scenario $y^L_{33} y_{32}<0$ and $y^L_{32} y_{33}>0$, the magnetic moment can reach the value $10^{-13}\mu_B$ for a LQ mass of 1.5 TeV, considering the upper bound on neutrino mass $\sum m_\nu\leq 0.26$ eV, as well as the other constraints considered in this work. In this work we claim for a fine-tuning of the LQs couplings to achieve a significantly enhanced neutrino magnetic moment while keeping the neutrino mass within the eV range. Furthermore, we leave the development of a symmetry that protects the neutrino mass for future investigations.

\section*{Acknowledgement}
I would like to thank Omar Miranda for his collaboration in the early stages of this project. This work has been supported by Secretaria de Ciencia, Humanidades, Tecnología e Innovación (SECIHTI) through Estancias Posdoctorales por México and Sistema Nacional de Investigadores (SNI) program under grant 629885.  This work was partially supported by CONAHCYT-México under grant A1-S-23238. Notice that the unpublished first version of the ArXiv manuscript 2508.00226v1, is different to the current article. Therefore, the second version should be considered as the one linked to the present article
\appendix
\section{Processes $\ell_i \to \ell_j \gamma$ and $a_\mu$}\label{appendixLVFdecay}
 
The contribution of the LQs $\slq$ and $\rlq$ to the LFV decay $\ell_i \to \ell_j\gamma$ arises at the one-loop level by Feynman diagrams similar to the presented in Fig.~\ref{fig:FeynmanDiag1} with the substitutions in the external leptons $\nu_{ \alpha,\beta} \to \ell_{i , j}$ and the replacement in the internal quarks $ \bar d_i \to \bar u_i $. Moreover, there are contributions from reducible diagrams, however they only give contributions to the monopole terms, which are canceled out with those arising from the irreducible diagrams due to gauge invariance. The decay amplitude $\ell_i \to \ell_j \gamma$ can be written as follows

\begin{equation}\label{eq:Amplitude2}
\mathcal{M} (l_i^- \to l_j^- \gamma) = - \frac{i e}{16 \pi^2} \epsilon_\mu^* (q) \bar u (p - q) \left ( A_L P_L + A_R P_R  \right ) \sigma^{\mu \nu} q_\nu u (p),
\end{equation}

where the form factors are given by
\begin{align}
A_L &= \frac{N_c \cos^2(\theta_{LQ})}{m_{\slq}^2}\sum_{k=1}^3 \Bigl[- \bigl(m_i y^R_{k i} y^{R*}_{k j} + m_j  y^{\prime L}_{k i} y^{\prime L*}_{k j} \bigr)\mathcal{H}\left(\frac{m_{u_k}^2}{m_{\slq}^2} \right)\nonumber \\
&+m_{u_k} y^{\prime L}_{k i} y^{R*}_{k j}\mathcal{I}\left(\frac{m_{u_k}^2}{m_{\slq}^2}\right)  \Bigr] +\left(\begin{array}{c} 
							m_{\slq}\to m_{\rlq}\\
							\cos(\theta_{LQ})\to \sin(\theta_{LQ}) 
							\end{array} \right),\\
A_R &= A_L(y^{\prime L}_{ki}\to y^R_{ki},y^R_{kj}\to y^{\prime L}_{kj}).
\end{align}
The loop functions are given by 
\begin{eqnarray}
\mathcal{I}(x) &=& \frac{7-8x+x^2+2(2+x)\ln x}{(1-x)^3},\\
\mathcal{H}(x)&=& \frac{1+4x-5x^2+2x(2+x)\ln x}{(1-x)^4}.
\end{eqnarray}
with $m_{u_k} = (m_u, m_c, m_t) $. Then, after averaging (summing) over polarizations of the initial (final) fermion and gauge boson, we use the respective two-body decay width formula to write down the branching ratio of $\ell_i^- \to \ell_j^- \gamma$ 
\begin{equation}\label{eq:Taudecay}
\mathcal{B} (l_i^- \to l_j^- \gamma) = \frac{\alpha_{ \text{em} } (m_i^2 - m_j^2)^3}{4 (4 \pi)^4 m_i^3 \Gamma_i} \left (|A_L|^2 + |A_R|^2 \right),
\end{equation}

with $\alpha_\text{em} = e^2/(4 \pi)$ and $\Gamma_i$ being the fine-structure constant and the total decay width of the charge lepton $l_i^-$ respectively. From Eq.~\eqref{eq:Amplitude2} we can subtract the expression for the muon magnetic moment induced by the scalar Leptoquarks
\begin{equation}\label{eq:muAMM}
\begin{split}
 a^{LQ}_\mu &= -\frac{N_c m_\mu \cos^2\theta_{LQ}}{6 (4\pi)^2 m^2_{\slq}}\sum_{k=1}^3 \Bigl[m_{u_k} \text{Re}(y^{\prime L}_{k 2} y^{R*}_{k 2})\mathcal{I}\left(\frac{m_{u_k}^2}{m_{\slq}^2}\right)\\
 &-m_\mu (| y^{\prime L}_{k 2}|^2 + |y^R_{k 2}|^2 )\mathcal{H}\left(\frac{m_{u_k}^2}{m_{\slq}^2}\right) \Bigr]+\left(\begin{array}{c} 
							m_{\slq}\to m_{\rlq}\\
							\cos(\theta_{LQ})\to \sin(\theta_{LQ}) 
							\end{array} \right)
\end{split}
\end{equation}

In the limit $m_{u_k} \ll m_{LQ}$, the $\mu$AMM can be reduced to 
\begin{equation}
 a^{LQ}_\mu = - \frac{N_c m_\mu m_k}{48 \pi^2 m_{\slq}^2} \text{Re} ( y^{\prime L}_{k2} y^{R*}_{k2})\sum_{k=1}^3 \left [4 \log \left ( \frac{m_{u_k}^2}{m_{\slq}^2} \right) + 7 \right ]+\left(\begin{array}{c} 
							m_{\slq}\to m_{\rlq}\\
							\cos(\theta_{LQ})\to \sin(\theta_{LQ}) 
							\end{array} \right) 
\end{equation}

\end{document}